\documentclass{nature}
\usepackage{graphicx,psfrag}
\usepackage{mathrsfs}
\usepackage{amsmath,amsfonts,amssymb}
\usepackage{multirow}
\usepackage{comment}
\usepackage{units}
\usepackage{enumerate}
\usepackage[normalem]{ulem} 
\usepackage{longtable}
\usepackage[rgb]{xcolor}
\usepackage{gensymb}
\usepackage{adjustbox}
\usepackage[none]{hyphenat}
\usepackage{multibib}
\usepackage{caption}
\newcites{New}{References}
\usepackage{lineno}
\usepackage{textcomp}

\newcommand{\mn}{{Mon. Not. R. Astron. Soc.}}
\newcommand{\mnras}{\mn}
\newcommand{\aj}{{"Astron. J."}}
\newcommand{\apj}{{Astrophys. J.}}
\newcommand{\apjl}{{Astrophys. J. Lett.}}
\newcommand{\apjs}{{Astrophys. J. Supp.}}

\newcommand{\aap}{{Astron. Astrophys.}}
\newcommand{\nat}{{Nature}}

\newcommand{\prd}{{Phys. Rev. D}}

\newcommand{\pasp}{{Pub. Ast. Soc. Pac.}}

\newcommand{\nar}{New Astronomy Reviews}
\newcommand{\bain}{Bulletin of the Astronomical Institutes of the Netherlands}

\title{The black hole low mass X-ray binary V404 Cygni is part of a wide hierarchical triple, and formed without a kick}
\author{Kevin B. Burdge$^{1,2}$ $^{*}$, Kareem El-Badry$^{3}$, Erin Kara$^{1,2}$, Claude Canizares$^{1,2}$, Deepto Chakrabarty$^{1,2}$,  Anna Frebel$^{1,2}$, Sarah C. Millholland$^{1,2}$, Saul Rappaport$^{1,2}$, Rob Simcoe$^{1,2}$, Andrew Vanderburg$^{1,2}$}

\begin{document}

\maketitle

\begin{affiliations}
 \item Department of Physics, Massachusetts Institute of Technology, Cambridge, MA 02139, USA
 \item Kavli Institute for Astrophysics and Space Research, Massachusetts Institute of Technology, Cambridge, MA 02139, USA
  \item Division of Physics, Mathematics and Astronomy, California Institute of Technology, Pasadena, CA, USA

\end{affiliations}

\begin{abstract}

Evidence suggests that when compact objects such as black holes and neutron stars form, they may receive a ``natal kick,'' where the stellar remnant gains momentum. Observational evidence for neutron star kicks is substantial\cite{Hobbs2005,Tauris2017}, yet limited for black hole natal kicks, and some proposed black hole formation scenarios result in very small kicks \cite{Fryer2001,Burrows2023}. Here, we report the discovery that the canonical black hole low-mass X-ray binary V404 Cygni is part of a wide hierarchical triple with a tertiary companion at least 3500 astronomical units away from the inner binary. Given the orbital configuration, the black hole likely received a sub-5 kilometer per second kick to have avoided unbinding the tertiary. This discovery reveals that at least some black holes form with nearly no natal kick. Furthermore, the tertiary in this system lends credence to evolutionary models of low-mass X-ray binaries involving a hierarchical triple structure\cite{Naoz2016}. Remarkably, the tertiary is evolved, indicating that the system formed 3-5 billion years ago, and that the black hole has removed at least half a solar mass of matter from its evolved secondary companion. During the event in which the black hole formed, it is likely that at least half of the mass of the black hole progenitor collapsed into the black hole; it may even have undergone a complete implosion, enabling the tertiary to remain loosely bound.

\end{abstract}

Black holes (BHs) are the stellar remnants of the most massive stars. Gravitational wave astronomy has raised key questions regarding the formation of these objects, including the role of natal kicks\cite{Callister2021} and dynamical interactions\cite{Naoz2016,OLeary2016,Rodriguez2016}. There are several proposed formation pathways for these stellar remnants, including some channels in which there is an implosion with no associated natal kicks or supernovae \cite{Fryer2001,Fryer1999,Mirabel2017}. The best constraints on BH kick physics come from X-ray binaries' Galactic orbits\cite{Mandel2016}, astrometric microlensing\cite{Andrews2022}, and orbital dynamics\cite{Shenar2022}. These constraints have largely ruled out the need for kicks of $>100 \,\rm km\,s^{-1}$ except for in a handful of systems with exceptional orbits\cite{Fragos2009,Brown2024}.

\begin{figure}
\centering
\includegraphics[width=1.0\textwidth]{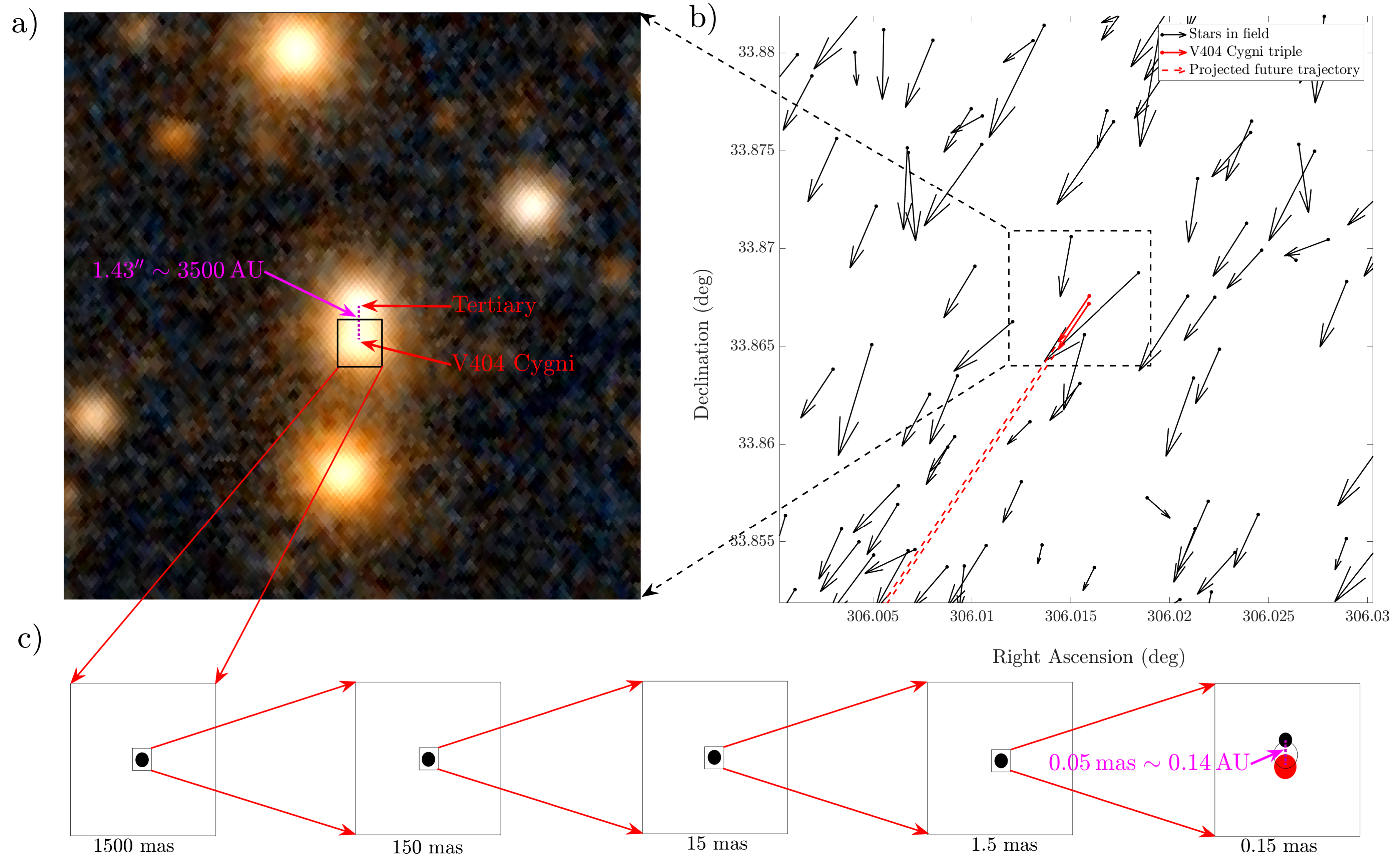}
\linespread{1.0}\selectfont{}
\caption{\textbf{a)}: A Pan-STARRS image of V404 Cygni and its companion, displayed using the Aladin interface, with V404 Cygni and the tertiary labeled in red and the separation of 1.43 arcseconds indicated in blue. We discovered the tertiary while viewing the source on Aladin and inspecting the Gaia astrometry on the interface, finding remarkable agreement in the measured proper motions of these two sources. \textbf{b)}: A plot of the positions and proper motion vectors of all stars in the field, with V404 Cygni and its tertiary indicated in red. \textbf{c)}: A zoom-in on the inner binary of V404 Cygni, illustrating the $2.5\times10^4$ ratio of the semi-major axes of the inner and outer orbits in the triple. }\label{fig1}
\end{figure}

\begin{figure}
\centering
\includegraphics[width=1.0\textwidth]{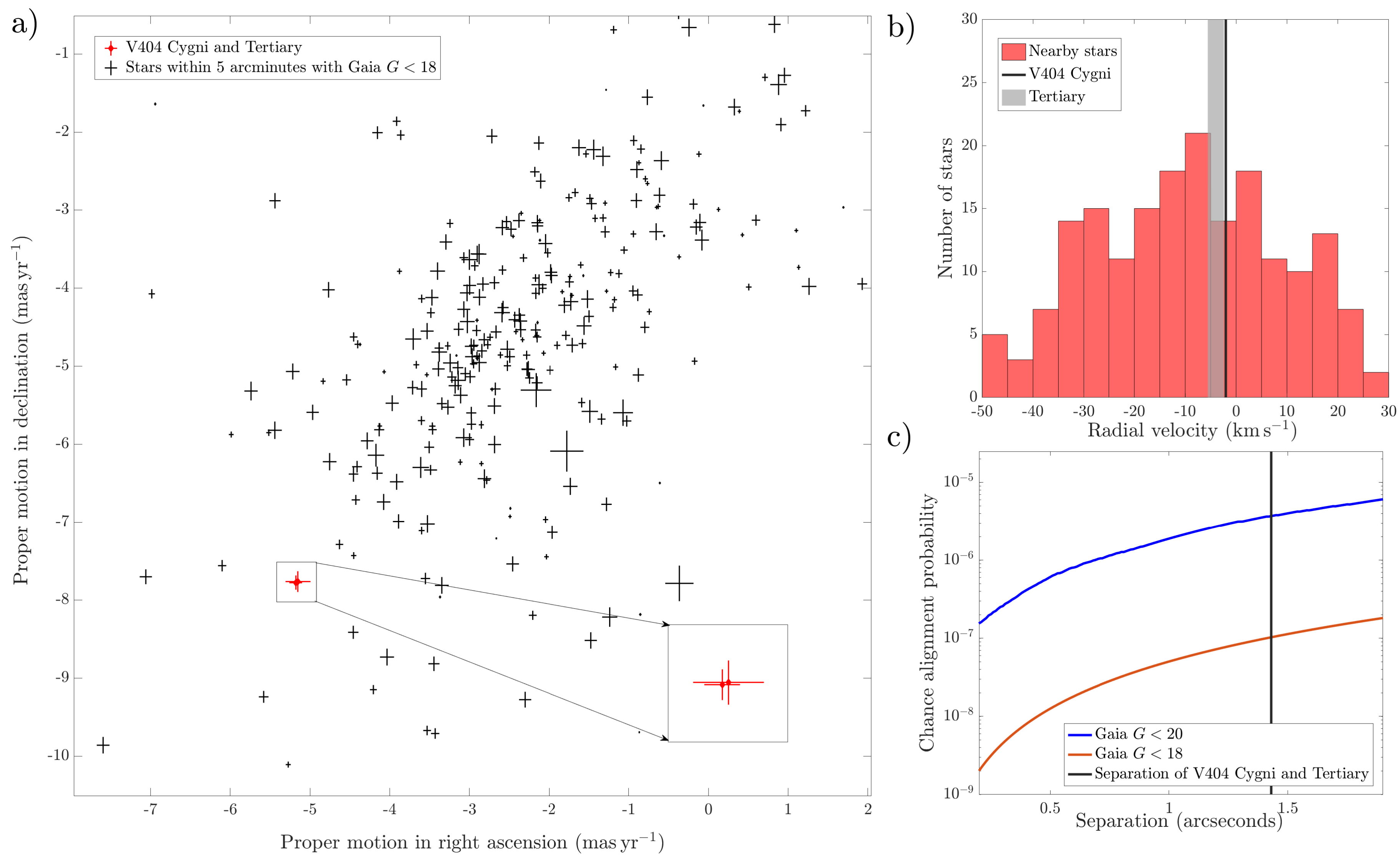}
\linespread{1.0}\selectfont{}
\caption{\textbf{a)}: A plot illustrating the proper motions of all Gaia stars brighter than 18th magnitude within 5 arcminutes of V404 Cygni, and the remarkable agreement between V404 Cygni and its tertiary (shown in red). \textbf{b)}: A histogram of Gaia-measured radial velocities of stars in the vicinity of V404 Cygni, with the systemic velocity of V404 Cygni shown as a black vertical line, and the measured radial velocity of the tertiary shown as the grey vertical line. \textbf{c)}: A plot illustrating the chance alignment probability as a function of separation for stars brighter than 18th magnitude (shown in red), and 20th magnitude (shown in blue). This probability accounts for both the agreement in proper motions, as well as radial velocity. }\label{fig2}
\end{figure}

We report the discovery that the black hole X-ray binary V404 Cygni, the first low-mass X-ray binary (LMXB) widely accepted to host a black hole (with a BH mass of $9^{+0.2}_{-0.6}\rm\, M_{\odot}$\cite{Khargharia2010}), is part of a wide hierarchical triple. Our serendipitous discovery stemmed from examining an optical image of V404 Cygni on the Aladin Lite tool, illustrated in panel a of Figure \ref{fig1}, and taking note that there is a nearby star just $1.43$ arcseconds from V404 Cygni with \emph{Gaia} proper motions matching those of V404 Cygni, as seen in panel b of Figure \ref{fig1}. We investigated the proper motions of nearby sources, and in panel a of Figure \ref{fig2}, show that such a chance agreement in proper motions is unlikely. In panel c of Figure \ref{fig2}, we quantify the probability of such an alignment occurring by chance, which we find is about $10^{-7}$ (Methods). While searching the literature on V404 Cygni, we found several works that commented on the nearby star, with most simply assuming that it was an interloper\cite{Udalski1991,Casares1991}. One notable exception was Maitra et al. 2017 \cite{Maitra2017}, which speculated ``in passing whether the blended star is truly unrelated to the V404 system'' due to the similar estimated distance and extinction of the blended star. V404 Cygni was noted as having a peculiar velocity in previous work\cite{Miller-Jones2009}, and this was attributed to a kick. As part of our astrometric analysis, we investigated nearby stars and found that the velocity of V404 Cygni is typical of stars in the vicinity (Methods).

Given the 1.43 arcseconds separation and estimated distance of $2.39\pm0.15\,\rm kpc$\cite{Miller-Jones2009}, we find that the wide tertiary companion is at least 3500 astronomical units (AU) away from the inner binary, which corresponds to a Keplerian velocity of just a few kilometers per second. As illustrated in panels a and c of Figure \ref{fig1}, this separation is approximately 90 times larger than the distance of Pluto to the Sun, and 25000 times larger than that of the inner V404 Cygni binary, which has a separation of just 0.14 AU, less than half the distance between the Sun and Mercury.

To further confirm this association, we analyzed archival VLT X-shooter spectroscopic observations (which targeted V404 Cygni, but contained a spatially resolved trace of the tertiary) and obtained additional follow-up GMOS spectroscopy of the source (Methods). We conducted a radial velocity analysis and found excellent agreement with the reported systemic velocity of V404 Cygni. As seen in the histogram of radial velocities of nearby stars shown in panel c of Figure \ref{fig2}, this is unlikely to have occurred by chance, further solidifying the association of the two components.

\begin{figure}
\centering
\includegraphics[width=1.0\textwidth]{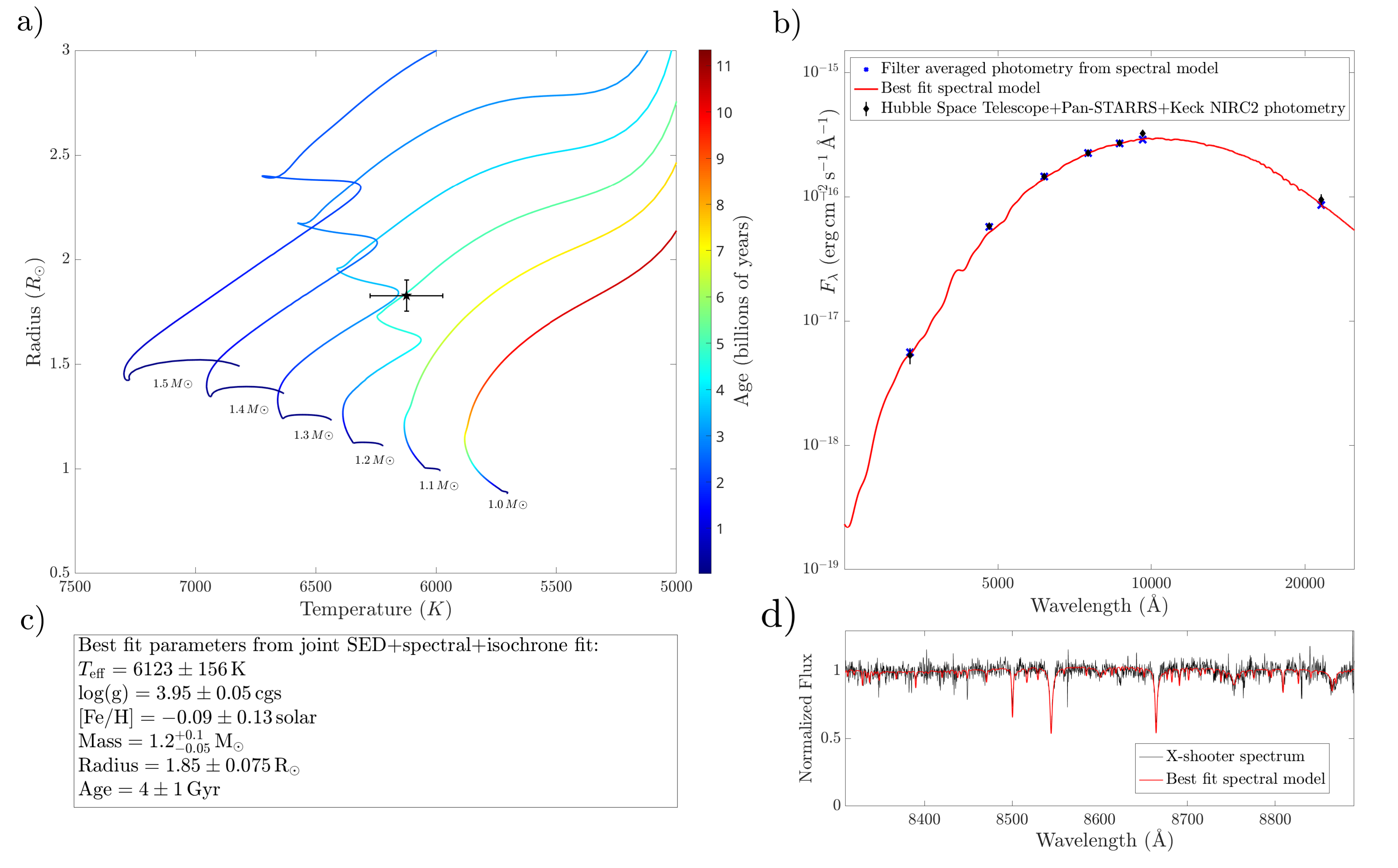}
\linespread{1.0}\selectfont{}
\caption{\textbf{a)}: MIST stellar isochrones, with the color bar indicating the age of stars evolving according to these isochrones. The black star with error bars indicates the position of the tertiary, which overlaps with tracks in the range of 1.2-1.3 solar masses at ages of around 3-5 billion years. \textbf{b)}: The spectral energy distribution of the tertiary in V404 Cygni (black diamonds), with the bluest point coming from Hubble Space Telescope ACS observations, the reddest point from Keck NIRC2 observations, and the remaining photometric measurements from Pan-STARRS images. The red curve shows our best-fit spectrum to this data, and the blue crosses indicate the filter-averaged flux values from this spectrum. \textbf{c)}: Our derived parameters as a result of the models illustrated in this figure. \textbf{d)}: The X-shooter spectrum of the Calcium triplet absorption features and other nearby absorption lines in the tertiary (black), and our best fit spectral model to it (red).  }\label{fig3}
\end{figure}

We fit the spectroscopic observations with model atmospheres, focusing on the region around the Hydrogen Alpha and Beta absorption lines in the GMOS spectra (Methods), and the region around the Calcium triplet lines in the X-shooter observations, shown in panel d of Figure \ref{fig3}. In addition to fitting the spectroscopic observations with model atmospheres, we also fit the broadband spectral energy distribution (SED), carefully extracting photometry by modeling the point-spread function in epochal Pan-STARRS images of the source due to the blending of the tertiary and V404 Cygni (Methods). Additionally, we use archival Hubble Space Telescope observations to obtain a measurement in the near ultraviolet at 330 nm, and archival observations with the Keck observatory's NIRC2 instrument to measure the near-infrared flux at $\sim 2$ microns. By jointly fitting the spectra and SED, we obtained constraints on the temperature, metallicity, and radius, with values reported in panel c of Figure \ref{fig3}.

One significant result in our modeling of the SED and spectra is that we find the tertiary in V404 Cygni has started to evolve off of the main sequence and is about twice its initial radius. By fitting the tertiary with MIST isochrones shown in panel a of Figure \ref{fig3}, we find that this constrains the system's age to about 3-5 gigayears, and the mass of the tertiary to around 1.2 solar masses (our full parameter estimates can be found in panel c of Figure \ref{fig3}). 

V404 Cygni's secondary, like the donors seen in some other BH LMXBs, exhibits enhanced lithium abundance, and this has been attributed to formation as a result of accretion processes onto the black hole, or in the supernova that formed it\cite{Martin1992}. Given our constraint on the age of the system, we can rule out that this lithium abundance is a result of recent formation; however, we inspected our X-shooter and GMOS spectra and concluded that at this time there is insufficient signal-to-noise and resolution to determine whether the tertiary has enhanced lithium.

To learn about the physical constraints imposed on the formation of the black hole, we simulated the dynamics of the triple with a range of configurations, accounting for the BH kick, mass lost during the BH formation, and the initial orbital periods of the secondary and tertiary, to investigate which BH formation scenarios could retain the loosely bound tertiary.

As seen in panel a of Figure \ref{fig4}, the only way the inner binary could have experienced a large kick and retained the tertiary, is if the tertiary started at a short orbital period, and was kicked into a highly eccentric orbit reaching $>3500$ AU at apastron. This scenario is unlikely, as one would need to fine-tune the kick to the inner system to be large, but just barely below the escape velocity of the system (Methods). We find that scenarios in which the tertiary started in a wide orbit, and the inner binary received a small kick of just a few kilometers per second are thus strongly favored.

When considering the inner binary, we simulated two scenarios. In one case, we allowed the secondary to start in a wide orbit between 100 AU and 300 AU. This scenario is viable for reproducing the current system, as we find that von Zeipel-Lidov-Kozai cycles could readily cause the secondary to migrate into its current 6.4-day orbit (Methods). As illustrated by the red histograms in panel b of Figure \ref{fig4}, we find that in this scenario, the BH kick was likely smaller than 3 kilometers per second and that the mass lost in the BH formation could have been up to about 10 solar masses, or about half the mass of the inner binary.

Alternatively, we consider a scenario where the secondary starts with an initial orbital period between 1 and 6 days (its current orbit is 6.4 days). These simulations are illustrated as the blue histograms in panel b of Figure \ref{fig4}. In this case, the inner binary's barycenter receives a significant Blaauw kick\cite{Blaauw1961} as a result of mass loss, even if the BH itself does not receive a kick. This results from matter being ejected in the BH progenitor's rest frame, causing it to be jettisoned out of the binary at the orbital velocity of the BH in the barycentric frame. At orbital periods of days, the BH orbits with a velocity of a few 10s of kilometers per second, and thus, as seen in panel b of Figure \ref{fig4}, this scenario strongly constrains any possible mass loss during the BH formation, because even if the BH does not receive a kick, the inner binary does as a consequence of the mass loss, ejecting the tertiary. One curiosity of this scenario is that in fine-tuned cases, one can achieve slightly larger BH kicks while keeping the tertiary bound, because the velocity imparted on the barycenter of the inner binary by mass loss can absorb the effect of the kick oriented in the opposite direction, resulting in a relatively low net kick to the inner binary, allowing the retention of the tertiary. Overall, this scenario still favors small kick velocities of less than 5 kilometers per second and does not allow for more than about a solar mass to be ejected. Thus, if the secondary started in a tight orbit of a few days, the most likely BH formation scenario is one in which there was a near-complete implosion of the progenitor star, with negligible mass loss. 

In either scenario, a BH formation event in which there is a complete implosion and no kick always results in the survival of the system. This would challenge current models for such a scenario, as this black hole has a mass of just $\sim 9\rm\, M_{\odot}$, which is a smaller BH mass than predicted for such an event\cite{Maitra2017}. We cannot rule out mass loss in the system, but we find it is improbable that more than half the BH progenitor's mass was suddenly ejected during its formation.

We consider it unlikely that the system formed dynamically in a dense environment such as a globular cluster or the Galactic center and was ejected, as the escape velocities of these environments exceed the orbital velocity of the tertiary by an order of magnitude (the young age of the tertiary also disfavors an origin in a globular cluster). We also find it unlikely that the system captured the tertiary in the field as a result of the low cross-section of such an interaction occurring.

The presence of a tertiary companion in one of the most well-known LMXBs supports theoretical work which has suggested that hierarchical triples may be key to forming BH LMXBs. Forming BH LMXBs via purely binary evolution has been theoretically challenging because of the large mass ratios involved, resulting in a common envelope event that proceeds to a merger rather than a successful ejection of the envelope. Thus, theoretical modeling of the formation of such systems has explored the possibility that wide tertiary companions helped the donor migrate into a tight orbit after the formation of the BH\cite{Naoz2016}. This is achieved through a gravitational interaction of the tertiary with the inner orbit, known as a von Zeipel-Lidov-Kozai cycle, in which the inner orbit cycles between an inclined and eccentric orbit. We note that models such as those presented in Naoz et al. 2016\cite{Naoz2016} predict tertiary companions in orbits at $\sim 10^4\rm AU$, which is consistent with what we observe in V404 Cygni. One plausible evolutionary scenario for V404 Cygni is that the inner binary began its life with a separation of $\sim 10^2\rm AU$, and over time Kozai-Lidov cycles drove this to a shorter orbit as a result of tidal dissipation and magnetic breaking draining orbital angular momentum from the inner binary during the highly eccentric phase. Finally, the inner binary hardened to an orbit of $<6.5$ days, too long for Roche-lobe overflow while the secondary was on the main-sequence, but as the secondary evolved, it overflowed its Roche-lobe. This is essentially the giant sub-channel described in Naoz et al. 2016 \cite{Naoz2016}. 

We note that searches for wide binary companions are, in general, highly incomplete. To be detected by {\it Gaia}, a companion must (a) be brighter than $G\sim 20.7$, and (b) must be separated from the inner binary by at least $\sim 1$\,arcsec (depending somewhat on flux ratio). For V404 Cygni, this corresponds to detection limits of $M \gtrsim 0.9\,M_{\odot}$ for main-sequence stars, and separations $s \gtrsim 2500$\,AU. The separation distribution of solar-type tertiaries peaks at 10-100 AU, with companions at 100-2500\,AU outnumbering those at $>2500$\,AU by a factor of two\cite{Raghavan2010, Tokovinin2014}. While the separation distribution of tertiaries to massive stars is quite uncertain, this suggests that companions too close to be detected by {\it Gaia} may be common and could have evaded detection thus far. Indeed, several BH LMXBs are thought to have unresolved companions, which have thus far been interpreted as chance alignments\cite{ArmasPadilla2019, MataSanchez2021, Hynes2002}. Proper motions for these companions have not yet been measured.

V404 Cygni is nearer and brighter in the optical than most other BH LMXBs. If the system were $\sim$50\% more distant, the companion would be blended with the inner binary, and {\it Gaia} would not have been able to measure its proper motion. If the companion were $\gtrsim 10\%$ more massive, it would already be a faint white dwarf below {\it Gaia}’s detection limit. If it were $\gtrsim 40\%$ less massive, it would be a main sequence star below the {\it Gaia} detection limit; if it were $\gtrsim 20\%$ less massive it would be too faint for {\it Gaia} to have measured a precise proper motion an establish the association with high confidence. These considerations all suggest that harder-to-detect tertiaries may well be hiding around other known BHs. It is quite possible that {\it most} BH LMXBs formed through triple evolution, and deeper searches around other BHs hold promise to detect them.

Evidence suggests that the spin axis of the BH in V404 Cygni is misaligned from the orbital plane due to the rapidly changing orientation of the jets in the system, and this has been attributed to a natal kick\cite{Miller-Jones2019}. However, the tertiary's presence largely rules out a natal kick. An alternative possibility is that the von Zeipel-Lidov-Kozai cycles induced this misalignment, as the inner orbit would have evolved through a range of inclinations and may have hardened at an inclination misaligned with the original orbit\cite{Liu2017, Su2021}. If the current spin of the black hole in the system was primarily inherited from the progenitor star, this could naturally lead to a misalignment of this spin axis with the current orbital plane. However, the evolved 1.2 solar mass tertiary implies the black hole has removed at least 0.5 solar masses from the 0.7 solar mass secondary—which was originally more massive than the tertiary, as it evolved first. If the black hole conservatively accreted this much mass, it could account for the large spin, but one would not expect a misalignment from the orbital plane. 

The tertiary companion of V404 Cygni has provided favorable evidence for the formation of at least some low-mass x-ray binaries in hierarchical triples. Moreover, it has provided one of the strongest empirical constraints on natal kicks in the formation of a black hole by indicating that the BH in V404 Cygni likely formed with a kick of less than five kilometers per second, demonstrating that at least some stellar mass black holes form without substantial natal kicks. We conclude by noting that our simulations strongly suggest that V404 Cygni's secondary either started in a wider orbit, and migrated in as a result of von Zeipel-Lidov-Kozai interactions, or if the secondary originated in a tight orbit, the $9\,M_{\odot}$ BH formed without ejecting more than a solar mass of matter--a near complete implosion.

\newpage

\begin{figure}
\centering
\includegraphics[width=1.0\textwidth]{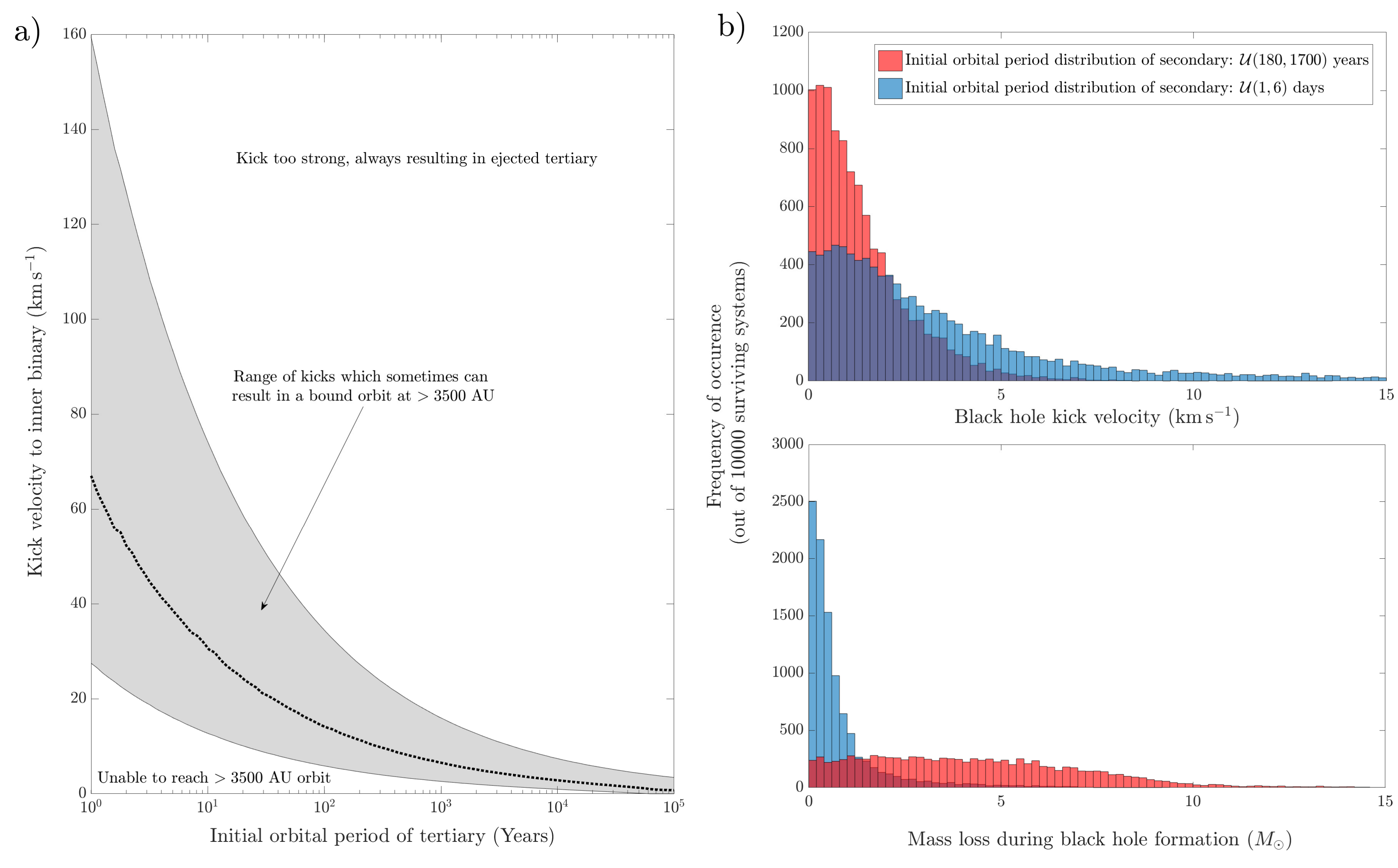}
\linespread{1.0}\selectfont{}
\caption{\textbf{a)}: An illustration of the range of possible kick velocities to the inner binary that could result in a bound tertiary at a separation greater than 3500 AU. In general, large kicks are only allowed if the tertiary originated at a very short orbital period, a scenario that is unlikely to retain the tertiary (Methods). \textbf{b)}: Histograms representing our simulations of black hole kicks and mass loss in the system. The red histograms represent a scenario in which the secondary originates at a wide range of orbital periods, whereas the blue histograms represent a scenario in which the secondary started at very short orbital periods.}
\label{fig4}
\end{figure}

\newpage

\begin{addendum}
 \item We dedicate this work to the memory of our dear friend, Tom Marsh. K.B.B. is a Pappalardo Postdoctoral Fellow in Physics at MIT and thanks the Pappalardo fellowship program for supporting his research. This research was supported by NSF grant AST-2307232. 

 \item[Competing Interests] The authors declare that they have no
competing financial interests.
 \item[Correspondence] Correspondence and requests for materials
should be addressed to K.B.B.~(email: kburdge@mit.edu).
\end{addendum}

\newpage

\begin{methods}

\subsection{Astrometric association}

We investigated the robustness of the astrometric association of the tertiary with the inner binary by analyzing a sample of all \emph{Gaia} sources within $30$ arcminutes of V404 Cygni. In \emph{Gaia} DR3, there are 80739 sources in this region of the sky. This corresponds to a source density of $0.0079$ sources per square arcsecond. As an initial check, we parsed these 80739 sources for objects with proper motion values that fall within 1 sigma of V404 Cygni's values and found only 21 such sources. When we account for the error bars of not just V404 Cygni's measured proper motions, but also of all other sources, we find that 1256 sources have 1 sigma confidence intervals that overlap the 1 sigma confidence interval of V404 Cygni. However, the mean magnitude in this sample is \emph{Gaia} G of 20.5, and thus it is dominated by sources with large uncertainties in proper motions, meaning that V404 Cygni's well-measured proper motion is consistent with over 1000 poorly measured proper motions in this region.

To construct the curve shown in panel c of Figure \ref{fig2}, which we define as the probability of finding a source within a given separation radius that has proper motions in RA and Dec consistent within 1 sigma, as well as a radial velocity consistent to within 1 sigma, we take the list of 80739 Gaia sources within 30 arcminutes, and further downselect to 39577 sources with a Gaia G magnitude greater than 20 (to construct the blue curve), and 10798 sources with a Gaia G magnitude greater than 18 (to construct the red curve). We follow the procedure outlined in El-Badry et al. 2018\cite{El-Badry2018} to determine whether two sources have consistent proper motions, accounting for possible orbital motion influencing the astrometric solution. We find that for stars with \emph{Gaia} $G>18$ (V404 Cygni's tertiary is 17.9), there are only 3 sources within 30 arcminutes that have proper motions consistent with V404 Cygni (this number ranges from 2-4 sources, as the number of sources consistent with the proper motions of V404 Cygni depends on the assumed separation, as this is used to compute possible variance in proper motions due to the orbital motion). In any case, this leads to small chance alignment probabilities. Extended Data Table 1 lists the \emph{Gaia} astrometric solution of V404 Cygni and its tertiary.

We also consider the radial velocity of the sources in computing the chance alignments. To do this, we selected all Gaia sources within 30 arcminutes of V404 Cygni with a Gaia radial velocity error less than 5 kilometers per second (this is comparable to our measured uncertainty on the RV of the tertiary). We used these sources to construct the histogram seen in panel b of Figure \ref{fig2}. From this distribution, we computed that there is approximately an 8.1 percent probability of a source with a well-measured RV being consistent with that of V404 Cygni to within 1 sigma, and we incorporated this information in our chance alignment probabilities illustrated in panel c of Figure \ref{fig2}.

To translate these numbers into formal chance alignment probabilities, we ask, what is the probability that such a source would fall within a circle of some radius (represented by the x-axis in panel c of Figure \ref{fig2}) around V404 Cygni. We use the vertical black line to denote the location of the actual tertiary, which is 1.43 arcseconds away from V404 Cygni. A 1.43 arcsecond region represents just 1/1584429 of the 30 arcminute region we queried for sources, and thus the probability of any individual source falling by chance into such a region is $6.3\times 10^{-7}$. Thus, if we consider the 3 sources with \emph{Gaia} $G>18$ that have proper motions consistent with V404 Cygni, accounting for possible orbital motion at a 1.43 arcsecond separation, the probability of one of those sources falling within 1.43 arcseconds of V404 Cygni is $1.26\times 10^{-6})$, and when we account for the probability of the radial velocity agreeing to within 1 sigma, this becomes $\sim10^{-7})$. 

While analyzing the astrometry of V404 Cygni and its tertiary, we used the astrometry of nearby stars to investigate the peculiar velocity reported in Miller-Jones et al. 2009 \cite{Miller-Jones2009}. Using stars within 1 degree of V404 Cygni, with similar distances (parallax values between 1/3 and 1/2), and radial velocity errors of less than $5\,\rm km\,s^{-1}$, we constructed the Toomre diagram shown in Extended Data Figure \ref{figToomre}, and found that the \emph{Gaia} astrometry and radial velocities of nearby stars are largely consistent with that of V404 Cygni.

\captionsetup[table]{name=Extended Data Table}

\begin{table}
\caption{Gaia astrometry of V404 Cygni and the Tertiary}\label{tab1}%
\scalebox{0.8}{ 
\begin{tabular}{@{}llllll@{}}
\scriptsize 
Object & RA (epoch 2016)  & Dec (epoch 2016) & $\omega$ & $\mu_{\rm RA}$ & $\mu_{\rm Dec}$\\

V404 Cygni    & $306.0159085525$   & $33.8671768648$  & $0.3024\pm0.0783$\footnotemark[1] & $-5.1775\pm0.0785$ & $-7.7776\pm0.0922$  \\
Tertiary    & $306.0159299085$   & $33.8675732739$  & $0.1423\pm0.1161$ & $-5.1500\pm0.1564$ & $-7.7647\pm0.1316$  \\

\end{tabular}
} \setcounter{table}{0} \caption{1 We note that the Gaia parallax measurement is much less precise than the radio parallax of $\omega=0.418\pm0.024$ reported in \cite{Miller-Jones2009}. The proper motions are consistent with what was measured in the radio.}
\end{table}

\begin{figure}
\centering
\renewcommand{\figurename}{Extended Data Figure}
\setcounter{figure}{0}  
\includegraphics[width=1.0\textwidth]{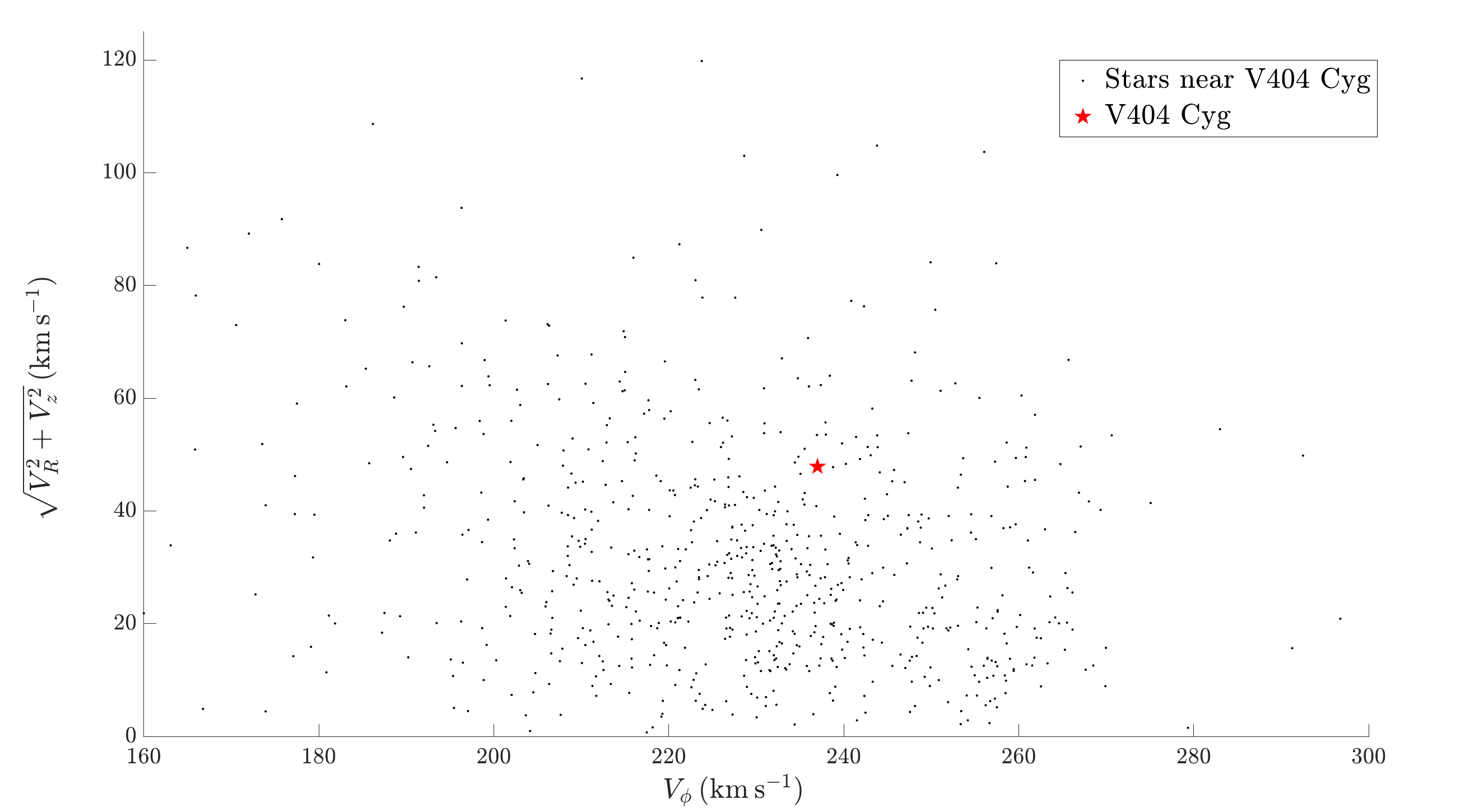}
\linespread{1.0}\selectfont{}
\caption{A Toomre diagram illustrating V404 Cygni's Galactic orbital velocity relative to stars within a degree of V404 Cygni. We find that V404 Cygni's velocity is largely consistent with nearby stars.}\label{figToomre}
\end{figure}

\subsection{Constraining kicks and mass loss}

To constrain the possible natal kicks the black hole could have received while still retaining the tertiary, we treated the system as a triple with component masses of $M_{BH}=8.5\,M_{\odot}$ as the post supernova black hole mass, $M_2=1.5\,M_{\odot}$ as the initial mass of the inner donor star, and $M_3=1.2\,M_{\odot}$ as the initial mass of the tertiary. 

We consider four free parameters in our simulations: the initial orbital period of the tertiary, the initial orbital period of the secondary, the amount of mass ejected in the supernova, and the kick that the black hole received. We marginalized over the kick angle and the orientation of the inner binary with respect to the outer tertiary's orbit, and in all cases we assumed initially circular orbits. In all cases, we assume initial orbits are circular. To determine whether the post-kick orbits remain bound, we follow the procedure outlined in Brandt \& Podsiadlowski (1995)\cite{Brandt1995}, which we outline here. We define a non-dimensional characteristic mass: 

\begin{equation}
\Tilde{m} = \frac{M_1 + M_2}{M'_1 + M'_2},
\end{equation}where $M_1$ and $M_2$ are the initial masses in the system, and $M'_1$ is the post-supernova mass of component that underwent mass loss. We also compute a dimensionless ratio of the magnitude of the kick velocity to orbital velocity:

\begin{equation}
\Tilde{v} = \frac{v_{\textrm{kick}}}{v_{\textrm{orb}}},
\end{equation}where $v_{\textrm{orb}}$ is the relative orbital velocity of the two components computed using Kepler's laws (in the case of the inner binary, the two components are the BH progenitor and the secondary, and in the case of the outer orbit, the components are the tertiary and the center of mass of the inner binary).

Following Brandt \& Podsiadlowski (1995)\cite{Brandt1995}, this yields a post-supernova energy in the new center of mass frame given by:
\begin{equation}
E' = - \frac{GM'_1M_2}{2a} \left[ 2 - \Tilde{m}(1 + 2 \Tilde{v} \cos \phi \cos \theta + \Tilde{v}^2) \right],
\end{equation}where $\phi$ and $\theta$ represent the polar and azimuthal kick angles, respectively. As observed in Brandt \& Podsiadlowski (1995)\cite{Brandt1995}, for the final system to remain bound, $E'$ must be negative, and the final semi-major axis is given by 
\begin{equation}
a' = - \frac{GM'_1M_2}{2E'}.
\end{equation}

In each simulation, we first use the BH kick and level of mass loss and consider the impact this has on the inner orbit's Barycentric velocity, $v_{\text{sys}}$. We then use this $v_{\text{sys}}$ as the input to a calculation to check whether the outer tertiary will remain bound to the inner binary. We compute $v_{\text{sys}}$ using the expression given in Brandt \& Podsiadlowski (1995)\cite{Brandt1995}:
\begin{equation}
v_{\text{sys}} = \frac{v_{\text{orb}}}{M'_1 + M_2} \left[ \left( \frac{\mu \Delta M_1}{M_1} \right)^2 - 2\frac{\mu \Delta M_1 M'_1}{M_1} \tilde{v} \cos \phi \cos \theta + (M'_1 \tilde{v})^2 \right]^{1/2},
\end{equation}where $\Delta M_1$ is the mass lost in the supernova, and $\mu$ is the reduced mass of the pre-supernova binary, defined as $\mu=\frac{M_1 M_2}{M1+M2}$.

In constructing the histograms shown in panel b of Figure \ref{fig4}, we consider two cases: one in which the inner companion started out in an orbit between 1 and 6 days (we use a uniform distribution of orbital periods between these two values), and another case in which it started out at an orbital separation between 100 and 300 AU (we also always enforce that the secondary must start out at a shorter orbital period than the tertiary). We use 100 AU as a lower bound to ensure that the von Zeipel-Lidov-Kozai timescales were short enough that the secondary would have migrated into its current orbit over the few Gyr age of V404 Cygni. We enforce the 300 AU upper bound because if the orbit were significantly wider, it would be dynamically unstable. In both scenarios, we drew from uniform distributions for the BH kick and mass loss during BH formation, with the BH kick being drawn from a uniform distribution between $0\,km\,s^{-1}$ and $300\,km\,s^{-1}$, and the mass loss being drawn from a uniform distribution ranging from $0\,M_{\odot}$ and $20\,M_{\odot}$. We consider ``surviving'' solutions to be ones where the secondary remains bound in an orbit with a semi-major axis less than the final semi-major axis of the tertiary and the tertiary remains bound, with a final semi-major axis greater than $3500\,\rm AU$. 

To construct panel a of Figure \ref{fig4}, which represents the range of possible kicks the inner binary could have experienced while retaining the tertiary as a function of the initial orbital period of the tertiary, we iterate over initial orbital periods of the tertiary and simply apply a range of systemic velocities to the inner binary, without assuming any mass loss, and investigate which solutions retain the tertiary in a bound orbit greater than $3500\,\rm AU$.

\begin{figure}
\centering
\includegraphics[width=1.0\textwidth]{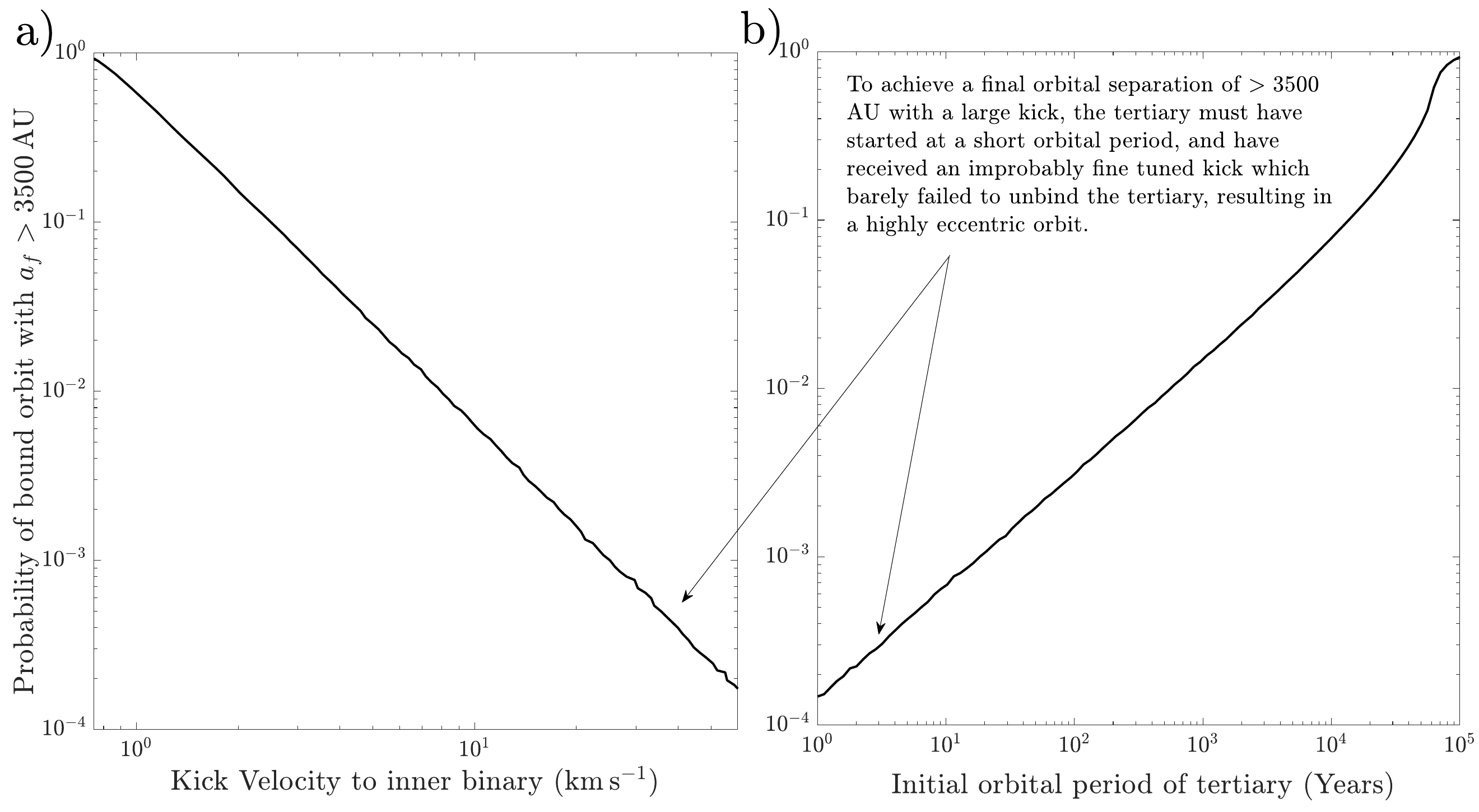}
\renewcommand{\figurename}{Extended Data Figure}
\linespread{1.0}\selectfont{}
\caption{\textbf{a)}: The probability of retaining a bound tertiary as a function of the kick velocity in the inner binary. In general, this probability diminishes at large kicks because these are only allowed when the tertiary originates in a tight orbit, and must be fine-tuned to be large enough to send it to a wide separation without unbinding it. \textbf{b)}: The probability of retaining a bound tertiary at greater than 3500 AU as a function of the initial orbital period of the tertiary. These probabilities are diminished for tertiaries that originate in tight orbits due to the large degree of fine-tuning required to send the tertiary to such a wide orbit without unbinding it.}
\label{figkick}
\end{figure}

\subsection{Spectroscopic Observations}

We searched the ESO archive for observations of V404 Cygni in which the companion may have serendipitously fallen inside the slit. This yielded a VLT/X-shooter spectrum obtained on 16 July 2015 (Program 295.D-5027; PI:Rahoui), between the main outburst and mini-outburst \cite{Casares2019}, when the black hole was in quiescence. To our knowledge, these data have not been published elsewhere. As seen in panel A of Extended Data Figure \ref{figxshooter}, two separate traces are clearly visible in the raw spectra, separated by about 1.4 arcsec, with one 5-10 times brighter than the other. Since V404 Cygni has no other comparably bright neighbors, there is little doubt that the fainter of the two traces corresponds to the companion. The seeing was about 1.3 arcsec (FWHM), meaning that the center of the companion's trace is separated from V404 Cygni by 2.5 times the ``$\sigma$'' of the seeing disk. Given that the companion is 5-10 times fainter than V404 Cygni in the VIS data, with its relative flux contribution increasing toward redder wavelengths, the companion was likely only partially in the slit. The data were taken with the 1.2 arcsec slit, yielding a typical resolution of 6500 in the VIS band.

The mid-exposure time is HJD 2457219.692, when the ephemeris of Casares et al. 2019\cite{Casares2019} predicts an RV of $-49.9\,\rm km\,s^{-1}$ for the donor. Twelve separate 148s exposures were obtained sequentially in nod-and-shuffle mode, for a total exposure time of 1776s in the VIS band. After inspecting the 12 exposures, we rejected one in which the companion was unusually faint, presumably because it fell farther outside the slit. We reduced and combined the other 11, for an effective exposure time of 1628s in the VIS band. We reduced the data using the ESO Reflex pipeline\cite{Freudling2013} with standard calibrations. This performs bias-subtraction, flat fielding, wavelength calibration using afternoon ThAr arcs, and order merging. We set the extraction window for the two sources manually using the \texttt{localize-slit-position} and \texttt{localize-slit-height} parameters. To minimize contamination from V404 Cygni, we extracted only the $\approx 50\%$ of the companion's trace on the far side of V404 Cygni. The fact that the companion's extracted spectrum shows no emission in H$\alpha$, where V404 Cygni shows a strong double-peaked emission line (see panels b and c of Extended Data Figure \ref{figxshooter}), suggests there is little contamination. 

\begin{figure}
\centering
\includegraphics[width=1.0\textwidth]{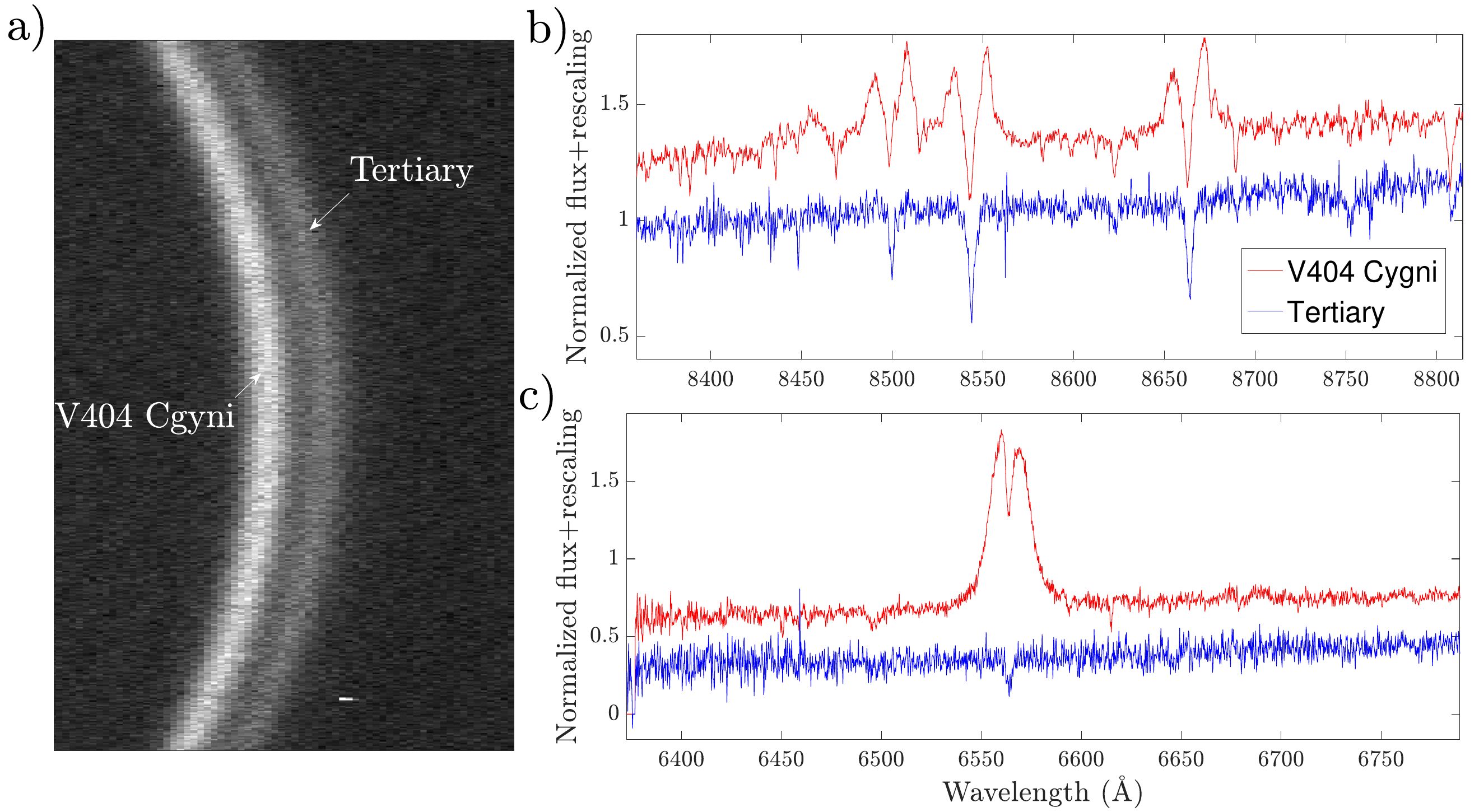}
\renewcommand{\figurename}{Extended Data Figure}
\linespread{1.0}\selectfont{}
\caption{\textbf{a)}: A region in the archival X-shooter spectrum of V404 Cygni, with the traces associated with the inner binary of V404 Cygni, and the tertiary both labeled. \textbf{b)}: The extracted X-shooter spectra around the calcium triplet region for V404 Cygni (shown in red), and the tertiary (shown in blue), illustrating that we were able to extract a spectrum of the tertiary with minimal contamination by V404 Cygni. \textbf{c)}: The spectrum of V404 Cygni (shown in red) and the tertiary (blue) around the H Alpha Hydrogen absorption line.}\label{figxshooter}
\end{figure}

We used the ``A'' and ``B'' telluric bands to perform a flexure correction in the wavelength solution, obtaining a $-12\rm \, km\, s^{-1}$ correction with ``A'' band, and a $-11.0\rm \, km\, s^{-1}$ with ``B'' band, and adopted a correction of $-11.5\rm \, km\, s^{-1}$, with a systematic uncertainty of $-1\rm \, km\, s^{-1}$. After applying these corrections, we converted the X-shooter spectrum to air wavelengths, and applied the Barycentric correction of $8.6\rm \, km\, s^{-1}$ in the file headers.

We also obtained an additional spectroscopic observation using Gemini Multi-Object Spectrographs (GMOS) on the 8.1-m Gemini North telescope on Mauna Kea, to determine whether there was any RV variability within the tertiary (program GN-2023B-DD-105). We used the 0.5-arcsecond slit and the R831\_G5302 grating, which provides a resolution of approximately 4400. Our spectrum covered a wavelength range from $4576 \,\text{\AA}$ to $6925 \,\text{\AA}$. While being slightly lower resolution than the X-shooter data, the spectrum provided a significantly higher signal-to-noise ratio (SNR) at bluer wavelengths.

\begin{figure}
\centering
\includegraphics[width=1.0\textwidth]{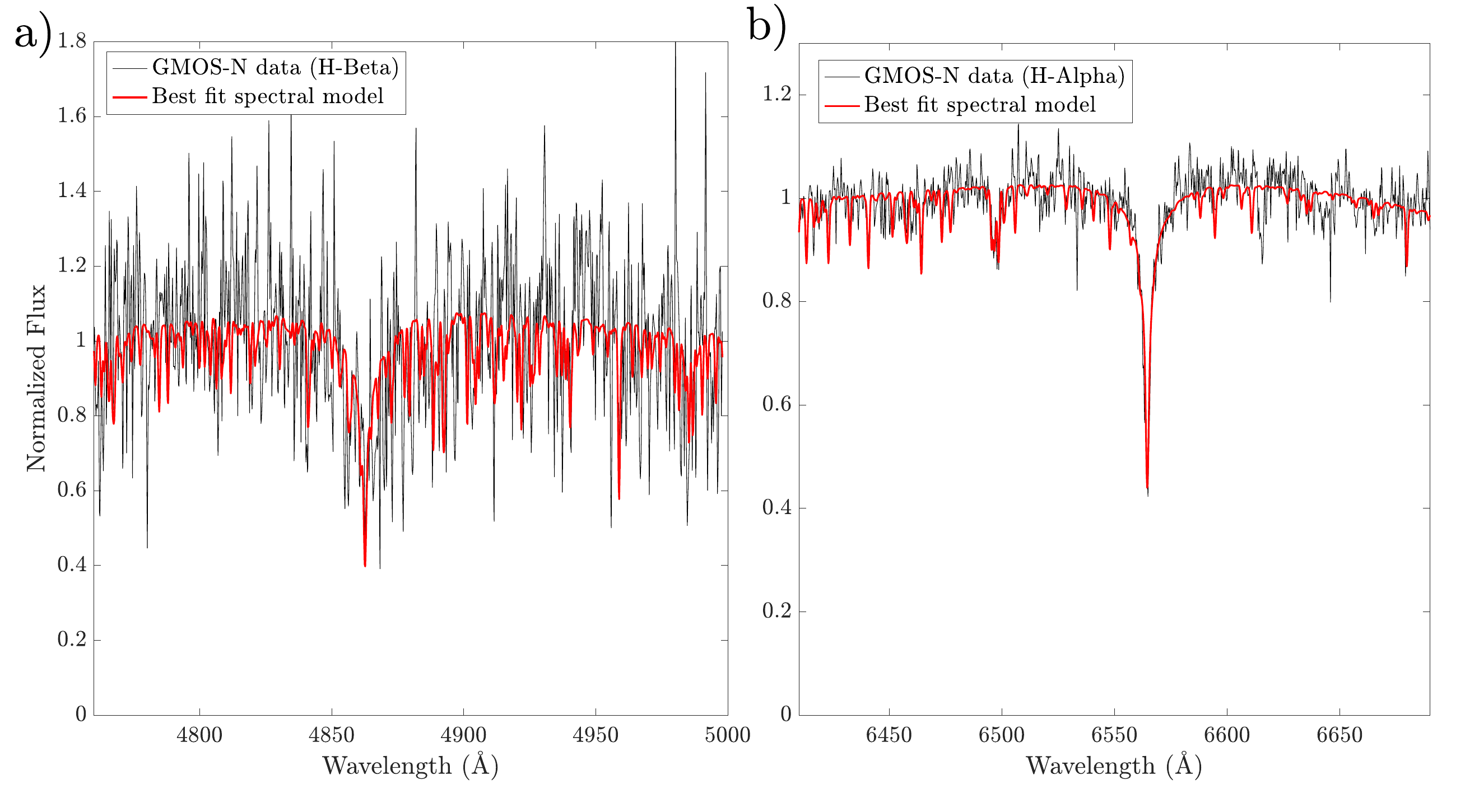}
\renewcommand{\figurename}{Extended Data Figure}
\linespread{1.0}\selectfont{}
\caption{\textbf{a)}: The Gemini GMOS-N spectrum of the H Beta absorption feature in the tertiary (black), and our best fit spectral model to it (red). \textbf{b)}: The Gemini GMOS-N spectrum of the H Alpha absorption feature in the tertiary (black), and our best fit spectral model to it (red).}\label{figgmos}
\end{figure}

We aligned the position angle of the slit perpendicular to the angle between V404 Cygni and the tertiary, effectively masking V404 Cygni. The observation was obtained in good seeing (FWHM $\approx 0.7$ arcsec), meaning that V404 Cygni was nearly 4 times the ``$\sigma$'' of the seeing disk from the nearest edge of the slit, and contamination is expected to be negligible. We used a 900 second exposure, yielding SNR of $\approx 30$ per pixel at 6600\,\AA, $\approx 10$ per pixel at 5600\,\AA, and $\approx 5$ per pixel at 5000\,\AA. We obtained a CuAr arc on-sky immediately after the science exposure. 

We reduced the GMOS data using \texttt{PypeIt}, which required manual construction of a new template for the R831 grating.  \texttt{PypeIt} performs bias and flat field correction, cosmic ray removal, wavelength calibration, sky subtraction, extraction of 1d spectra, and heliocentric RV corrections. Like the X-shooter data, we checked the wavelength solution using the telluric bands, and in this case found that no correction was needed. 

\subsection{Radial Velocity Measurements}

We used the X-shooter spectra to measure the radial velocities (RVs) of both components, as well as perform atmospheric fits in conjunction with modeling the spectral energy distribution (SED). To measure RVs, we used the ultranest kernel density estimator to fit an atmospheric model constructed using the PHOENIX stellar atmosphere library\cite{Husser2013}. We selected a region between $8480 \rm \,\text{\AA}$ and $8700 \rm \,\text{\AA}$ for the fit, as this wavelength range contains the calcium triplet lines, which provide a strong RV signal in the x-shooter spectrum. We allowed the RV to be a free parameter, and fixed the other parameters in this atmospheric model to an effective temperature of $T_{eff}=6100\,\rm K$, $logg=3.95\,\rm cgs$, and $[Fe/H]=0.0$ based on our joint atmospheric+SED fit. Our fit yielded an RV measurement of $-4.5\pm2.5\pm1\rm\, km\,s^{-1}$ for the tertiary, in excellent agreement with the $-2.0\pm0.4\rm\, km\,s^{-1}$ systemic velocity of the inner binary reported in Casares et al. 2019\cite{Casares2019}.

We also conducted a radial velocity analysis of the GMOS spectra, fitting the region between $6400\rm \,\text{\AA}$ and $6700\rm \,\text{\AA}$ to capture the H alpha absorption line and nearby features, and the region between $4750\rm \,\text{\AA}$ and $5000\rm \,\text{\AA}$ for the H beta absorption feature. We find that this analysis yields a radial velocity of $-4.1\pm1.9\rm\, km\,s^{-1}$, consistent with that of the X-shooter data, and of V404 Cygni's systemic velocity. We note that the orbital motion of the tertiary is on the order of a few kilometers per second, and may also introduce some difference between the systemic velocities of the two components at this level.

\subsection{Spectral energy distribution}

To model the spectral energy distribution of the tertiary in V404 Cygni, we used a combination of observations obtained with the Panoramic Survey Telescope and Rapid Response System (Pan-STARRS)\cite{Chambers2016}, the Advanced Camera for Surveys (ACS)\cite{Sirianni2005} aboard the Hubble Space Telescope (HST), and near-infrared camera (NIRC2) on the W.M. Keck Observatory\cite{Matthews1994}. 

Because of the overlapping point spread function of V404 Cygni's inner binary and the tertiary in the Pan-STARRS images, we performed our own point spread function photometry of the source by constructing a routine using the photutils package to iteratively run over the epochal Pan-STARRs images. We used a small cutout region, illustrated in panel b) of Extended Data Figure \ref{figPS} for our analysis, constructing a PSF model using the nearby bright stars near the top and bottom of this cutout. After constructing the PSF model, we use photutils to identify bright sources in each image and fit for the position of the PSF, as seen in panel c) of Extended Data Figure \ref{figPS}, and we found that this yielded good quality photometry, leaving only small residuals, as seen in panel d) of Extended Data Figure \ref{figPS}. We note that initially when we performed this analysis on the stacked Pan-STARRS images, we were unable to eliminate substantial residuals, and discovered after inspecting the epochal images that there was a nearby focal plane artifact, in the shape of a heart, which may have been contaminating some of the stacked images. One example of this artifact is illustrated in panel a) of Extended Data Figure \ref{figPS}, and we found that its position regularly changed at different pointings, sometimes contaminating V404 Cygni or our comparison stars, leading to a variable PSF model over the stacked images. To quantify the uncertainties in the apparent magnitude of the V404 Cygni tertiary, we computed the standard deviation of the estimated apparent magnitude across all epochal Pan-STARRS images for which we were able to perform photometry (in each filter). In addition to using the photometric scatter across epochal images to estimate the measurement error, we compared the inferred apparent magnitude using the star just to the north of the triple, and the one just to the south of the triple, to quantify the systematic error in each filter. We note that we used the Pan-STARRS1 reference catalog of apparent magnitudes for our two comparison stars as a basis for calibrating the photometry.

We used an archival HST observation (Proposal ID: 9686, PI: Hynes) to extract flux in the ACS/HRC F330W filter. These observations consist of two 600s exposures. As seen in panel a of Extended Data Figure \ref{figHST}, the tertiary is visible and well resolved from V404 Cygni in these images. We used the ACSTools python package to compute the flux of the tertiary, using a 0.2 arcsecond aperture and applying the appropriate corrections to the flux to account for the portion of the PSF not encompassed by this aperture.

We performed infrared photometry using a single archival NIRC2 image of V404 Cygni, shown in panel b of Extended Data Figure \ref{figHST}. We used a 0.5 arcsecond aperture to extract flux for the tertiary, as well as a comparison star just to the north of the tertiary. We used the 2MASS reference magnitude of this reference star to calibrate our flux, though we elected to apply a systematic error of ten percent to our photometry due to the difference in filters in the 2MASS system (2MASS $K_s$), and the $K_p$ filter used in the NIRC2 observation. We elected not to perform more in-depth photometry on the full archive of hundreds of images because we found that this photometric measurement had little overall impact on our results.

\begin{figure}
\centering
\includegraphics[width=1.0\textwidth]{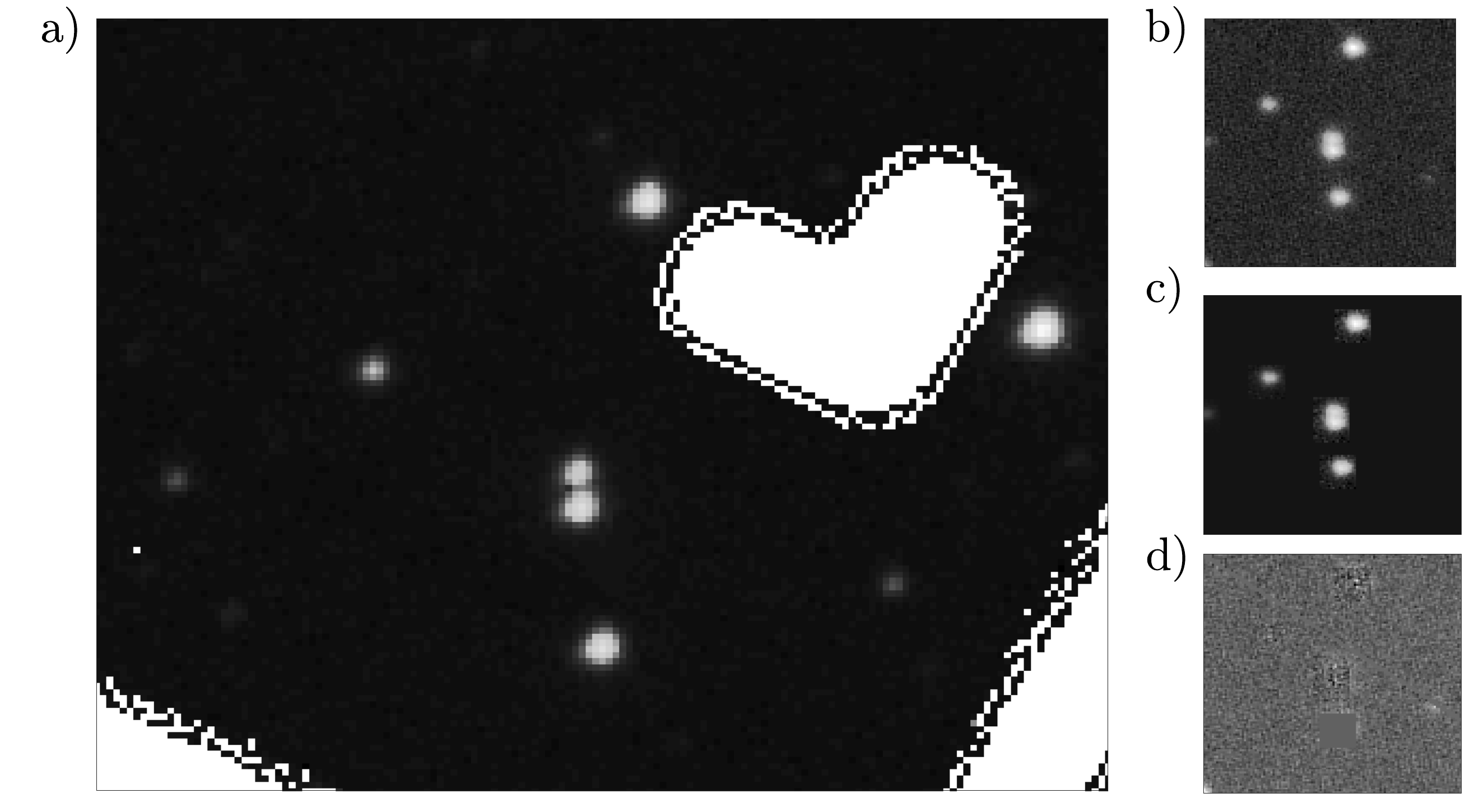}
\renewcommand{\figurename}{Extended Data Figure}
\linespread{1.0}\selectfont{}
\caption{\textbf{a)}: An example of a heart-shaped artifact that contaminates many of the Pan-STARRS images of V404 Cygni. We believe this artifact degraded the quality of the stacked PS images by occasionally occluding either V404 Cygni or comparison stars, resulting in a non-uniform PSF across the stacked images. This is why we ultimately decided to use epochal images as the basis of our photometry. \textbf{b)}: An epochal Pan-STARRS r-band image of V404 Cygni. We used epochal images such as this to extract fluxes for our spectral energy distribution analysis. \textbf{c)}: An image illustrating the point spread function (PSF) model we applied to the image to extract photometry for V404 Cygni's tertiary. \textbf{d)}: A cutout illustrating the residuals after we extract PSF photometry, with the PSF model constructed using the bottom star in the image (hence the zero residuals around this star). }\label{figPS}
\end{figure}

\subsection{Joint SED, Spectroscopic, and Isochrone analysis}

We performed an analysis in which we fit the SED with model atmospheres, synthesizing fluxes using the pyphot tool. In addition to the SED measurement, the only other measurement we included in this analysis was the estimated radio parallax of $0.418\pm0.024$ to constrain the distance (and appropriately account for the uncertainty in distance in estimating the radius), and the reddening value of $E(g-r)=1.27$ reported in Green et al. 2019\cite{Green2019}. As with the analysis of the spectra, we used the ultranest kernel density estimate to perform this fit. We fit for four free parameters, the distance, effective temperature, metallicity, and radius of the star. This analysis of the SED alone yielded the following parameters: $T_{eff}=6080\pm 86\,K$, $r=1.882\pm0.047\,R_{\odot}$, $d=2.42\pm0.14\,kpc$, and $[Fe/H]=-0.21\pm0.28$. We note that these estimates depend strongly on the assumed reddening, but are broadly consistent with what we estimate from the spectra.

\begin{figure}
\centering
\includegraphics[width=1.0\textwidth]{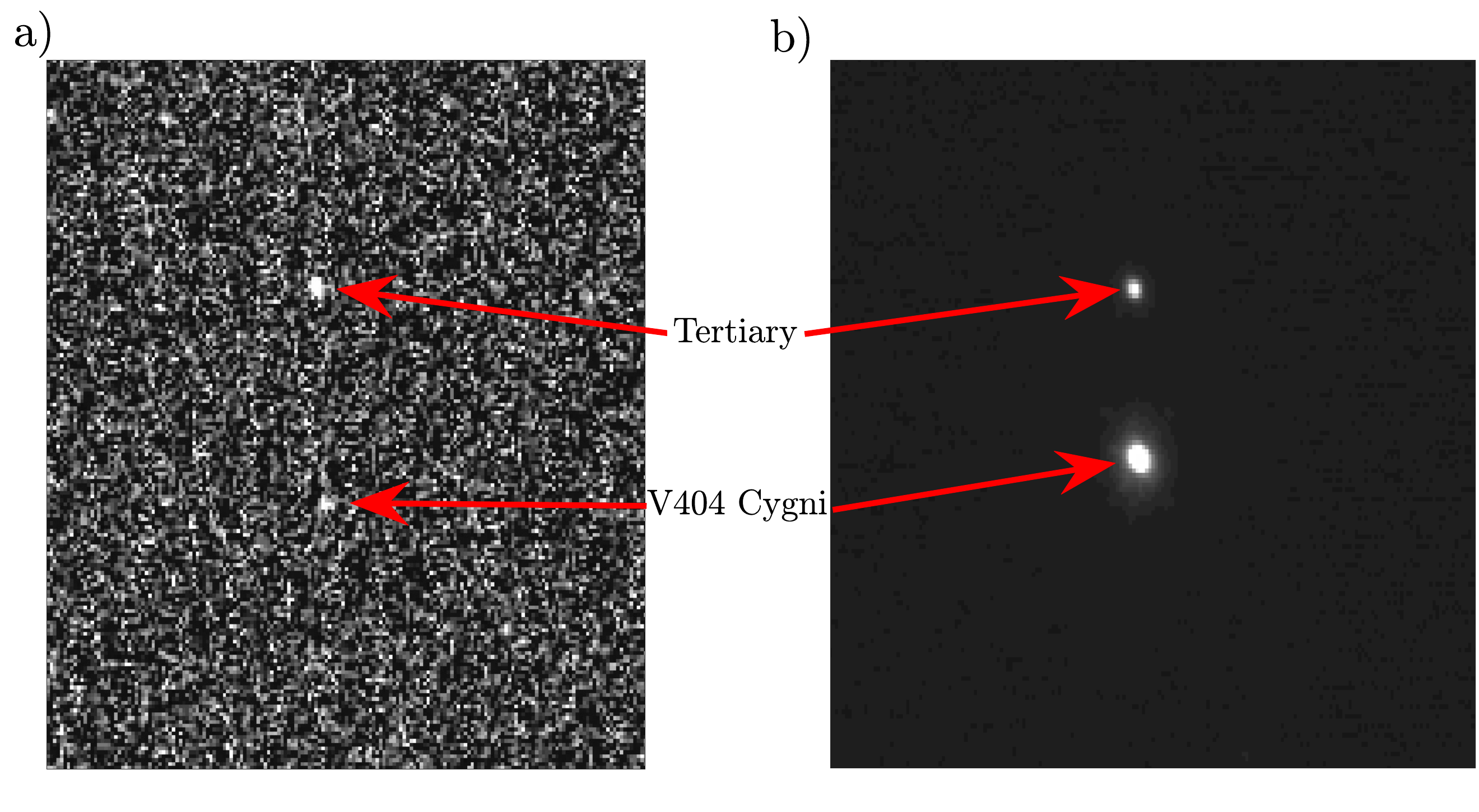}
\renewcommand{\figurename}{Extended Data Figure}
\linespread{1.0}\selectfont{}
\caption{\textbf{a)} A Hubble Space Telescope ACS/HRC F330W image of V404 Cygni, with both objects labelled. We used this archival image to extract flux in the F330W filter, which was included in our analysis of the spectral energy distribution of the object. \textbf{b)} A Keck NIRC2 adaptive optics image obtained in the $K_{p}$ filter, used in our analysis of the spectral energy distribution of the object.}\label{figHST}
\end{figure}

\end{methods}

\end{document}